\documentclass[11pt]{article}

\usepackage[a4paper,margin=1in]{geometry}
\usepackage[T1]{fontenc}
\usepackage[utf8]{inputenc}
\usepackage{lmodern}
\usepackage{amsmath,amssymb,amsfonts,bm}
\usepackage{physics}
\usepackage{graphicx}
\usepackage{tikz}
\usetikzlibrary{arrows.meta,positioning,fit,calc,shapes.geometric}
\usepackage{booktabs}
\usepackage{microtype}
\usepackage{hyperref}
\usepackage[numbers,sort&compress]{natbib}
\usepackage{enumitem}
\usepackage{xcolor}

\hypersetup{colorlinks=true,linkcolor=blue!60!black,citecolor=blue!60!black,urlcolor=blue!60!black}

\title{Fisher--Informational Time: A Causal-Geometric Framework for Emergent Clock Time}
\author{Juan Sumaya-Mart\'inez\thanks{Corresponding author: \texttt{jsm@uaemex.mx}}\\
\small Facultad de Ciencias, Universidad Aut\'onoma del Estado de M\'exico, Toluca, M\'exico}
\date{Version 2.0 --- \today}

\newcommand{\LF}{\Lambda_F}
\newcommand{\LQ}{\Lambda_Q}
\newcommand{\gF}{g^{(F)}}

\newcommand{\FS}{\mathrm{FS}}
\newcommand{\CRB}{Cram\'er--Rao bound}

\begin{document}
\maketitle

\begin{abstract}
We develop a Fisher--informational reformulation of physical time in which clock time is not regarded as a fundamental ontological substance, but as an emergent calibration of causally ordered distinguishability among physical states. The operational starting point is that clocks do not measure time itself; rather, they instantiate reproducible physical processes whose distinguishable states are correlated with other events. We introduce a causal--informational parameter, denoted by $\LF$, defined as an accumulated Fisher-geometric distance along a causally admissible trajectory in state space. In classical statistical systems, this parameter is generated by the Fisher information metric; in quantum systems, the corresponding construction is associated with the quantum Fisher information, the Bures metric, and the Fubini--Study geometry of projective Hilbert space. Version 2 of this manuscript strengthens the mathematical formulation, distinguishes model-dependent Fisher information from quantum Fisher information, clarifies the reparameterization of Schr\"odinger dynamics, and adds explicit examples for a qubit clock, an exponential decay process, and a Fisher characterization of clock quality. The proposal is positioned relative to relational time, the Page--Wootters mechanism, thermal time, quantum speed-limit relations, information geometry, and the problem of time in quantum gravity. We do not claim that relational or emergent time is new. The specific contribution is the use of Fisher distinguishability as an operational precursor from which ordinary clock time can be reconstructed. In this sense, the central statement of the paper is: \emph{time is not measured by clocks; clock time is reconstructed from the Fisher distinguishability accumulated along causally ordered physical changes}.
\end{abstract}


\section{Introduction}
Time is among the most useful variables in physics and among the most conceptually difficult. In Newtonian mechanics, time is introduced as an external universal parameter. Dynamical variables are written as
\begin{equation}
q=q(t),
\end{equation}
and the variable $t$ is assumed to flow uniformly and independently of the system. Special and general relativity weaken this interpretation: simultaneity becomes frame-dependent, proper time is path-dependent, and gravitational fields affect clock rates. Quantum mechanics introduces a further asymmetry: time is normally not represented as a self-adjoint observable on the same footing as position or momentum, but appears as an external parameter in the Schr\"odinger equation.

This tension becomes particularly severe in quantum gravity. In canonical quantum gravity, the absence or ambiguity of an external time parameter leads to the well-known problem of time, discussed extensively by Isham, Kucha\v{r}, Rovelli, Anderson, and others \cite{Isham1993,Kuchar2011,Rovelli1990,Anderson2012}. Relational approaches, including the Page--Wootters mechanism, recover apparent evolution from correlations between a clock subsystem and the rest of the universe \cite{PageWootters1983,Wootters1984,MarlettoVedral2017}. The thermal time hypothesis of Connes and Rovelli suggests that physical time-flow may be determined by the thermodynamic state of a system rather than by an external parameter \cite{ConnesRovelli1994,Rovelli2009ThermalTime}. Modern quantum-information approaches to clocks and time dilation also emphasize that clock readings are physical correlations rather than primitive temporal facts \cite{Smith2020QuantumClocks,Giovannetti2006QuantumMetrology,Pezze2018QuantumMetrology}.

The present work develops a complementary idea. It begins with a simple operational observation: a clock does not measure time as an independent substance. A pendulum counts oscillations, a quartz clock counts mechanical vibrations, and an atomic clock counts quantum transitions. In each case, what is physically realized is a reproducible sequence of distinguishable states. These states are then correlated with other physical events. Thus, a clock does not provide access to time itself; it provides a stable physical process that can be used to order events.

This paper proposes that the deeper object is not time, but causally ordered distinguishability. We introduce an accumulated Fisher-geometric parameter $\LF$, intended to play the role of an informational precursor to clock time. The core proposal is:
\begin{quote}
\emph{Clock time is reconstructed from the Fisher distinguishability accumulated along causally ordered physical changes.}
\end{quote}
The purpose is not to discard the empirical success of ordinary time. Rather, the goal is to reinterpret time as an emergent calibration of physical distinguishability. Ordinary time remains valid whenever stable clocks exist and when their distinguishable state sequences can be mapped monotonically to the processes under study.

\section{Operational Motivation: What Does a Clock Measure?}
A clock is a physical system whose states can be reliably counted, compared, and correlated with other states. Operationally, a clock establishes a relation of the form
\begin{equation}
E_A \longleftrightarrow N\;\text{cycles of a reference process}\longleftrightarrow E_B,
\end{equation}
where $E_A$ and $E_B$ are events and $N$ is an integer or real-valued count associated with the clock. The quantity called elapsed time is reconstructed from this correlation.

This motivates a distinction:
\begin{equation}
\text{objective clock correlations} \neq \text{ontological time substance}.
\end{equation}
The correlations are physical and reproducible. However, the inference that there exists a substance called time flowing independently of physical processes is not required.

The analogy with temperature is useful. Temperature is not a microscopic substance, but it is a well-defined emergent variable associated with statistical properties of many degrees of freedom. Similarly, time may be an emergent variable associated with ordered, reproducible, and distinguishable physical change. This paper is not a denial of measured durations, but a reinterpretation of what is actually measured.

\section{Central Hypothesis}
We formulate the central hypothesis as follows:
\begin{quote}
\textbf{Fisher--informational time hypothesis.} Physical time is not fundamental. It is an emergent calibration of the accumulated Fisher distinguishability of causally connected physical states.
\end{quote}
Instead of taking
\begin{equation}
q=q(t)
\end{equation}
as fundamental, we propose a deeper parameterization
\begin{equation}
q=q(\LF),
\end{equation}
where $\LF$ is an informational proper parameter. The ordinary time variable $t$ appears only when a clock subsystem $C$ is selected and its sequence of distinguishable states is used to calibrate other physical changes:
\begin{equation}
t_C=f_C\!\left(\LF^{(C)}\right).
\end{equation}
Thus, time is not abolished; it is reconstructed.

\section{Postulates}
\subsection{Postulate I: Physical distinguishability}
Only physically distinguishable changes carry operational content. If two states cannot be distinguished by any physically allowed measurement, assigning a temporal separation between them has no operational meaning. In statistical language, distinguishability is encoded in the response of probability distributions to parameter changes.

\subsection{Postulate II: Causal admissibility}
Physical changes must be causally ordered. We write
\begin{equation}
A\prec B
\end{equation}
to mean that event or state $A$ can causally precede or influence event or state $B$. The primitive structure is therefore not necessarily $t_A<t_B$, but the causal relation $A\prec B$.

\subsection{Postulate III: Fisher-geometric separation}
For a family of states described by probability distributions $p(x|\theta)$, infinitesimal distinguishability is quantified by the Fisher information metric
\begin{equation}
\gF_{ij}(\theta)=\int dx\,p(x|\theta)\,
\partial_i\ln p(x|\theta)\,\partial_j\ln p(x|\theta),
\end{equation}
where $\partial_i\equiv\partial/\partial\theta^i$. This metric induces an informational line element
\begin{equation}
d\LF^2=\gF_{ij}(\theta)d\theta^i d\theta^j.
\end{equation}

\subsection{Postulate IV: Emergent clock time}
Clock time emerges when a physical reference process provides a monotonic calibration of $\LF$. If a clock trajectory has accumulated Fisher distance $\LF^{(C)}$, then clock time is a conventional but reproducible mapping
\begin{equation}
t_C=f_C\!\left(\LF^{(C)}\right).
\end{equation}
Different clocks may realize different mappings, but agreement emerges under synchronization and calibration procedures.

\section{Fisher Information as Physical Distinguishability}
Fisher information was introduced by Fisher in statistical estimation theory \cite{Fisher1925}. For an unbiased estimator $\hat\theta$, the \CRB{} states that, under regularity assumptions,
\begin{equation}
\mathrm{Var}(\hat\theta)\geq \frac{1}{I(\theta)}.
\end{equation}
Thus, Fisher information quantifies the local statistical distinguishability of nearby parameter values. If a small change in $\theta$ produces a large change in the observable probability distribution, then $I(\theta)$ is large and the parameter can be estimated precisely.

In the present framework, this statistical meaning is reinterpreted dynamically. A physical change is significant not because an external time variable has advanced, but because the state has become more distinguishable. The accumulated distinguishability along a path $\gamma:\theta^i(\lambda)$ is
\begin{equation}
\LF[\gamma]=\int_\gamma \sqrt{\gF_{ij}(\theta)\frac{d\theta^i}{d\lambda}\frac{d\theta^j}{d\lambda}}\,d\lambda,
\end{equation}
where $\lambda$ is an arbitrary path parameter. The physical content is not in $\lambda$, but in the invariant length $\LF$.

This is analogous to arc length in geometry. A curve can be parameterized in many ways, but its length is invariant under reparameterization. Likewise, physical evolution may be parameterized by time, phase, redshift, entropy, or another monotonic variable, while the accumulated distinguishability provides an invariant measure of change relative to the chosen statistical description.

\subsection{Classical Fisher information versus quantum Fisher information}
An important qualification is required. Classical Fisher information depends on the measurement model $p(x|\theta)$. It is therefore operational but not measurement-independent. If different measurements are performed, different classical Fisher metrics may result.

The quantum Fisher information (QFI) is more fundamental for quantum systems because it optimizes distinguishability over all measurements allowed by quantum mechanics \cite{BraunsteinCaves1994,Paris2009,Helstrom1976,Holevo2011}. Thus, in the hierarchy proposed here,
\begin{equation}
\text{QFI} \quad \longrightarrow \quad \text{optimal distinguishability},
\end{equation}
whereas classical Fisher information is the experimentally realized distinguishability associated with a specified measurement protocol. This distinction is essential if $\LF$ is to be interpreted as more than a data-analysis tool.

\section{Quantum Extension: QFI, Bures Metric, and Fubini--Study Geometry}
In quantum mechanics, states are represented by density operators $\rho(\theta)$ or, for pure states, by rays in Hilbert space. Wootters introduced a statistical distance for quantum states \cite{Wootters1981}, and Braunstein and Caves showed that statistical distinguishability leads naturally to a Riemannian geometry on the space of quantum states through the QFI \cite{BraunsteinCaves1994}. A geometric perspective on quantum parameter estimation is reviewed in Ref.~\cite{SidhuKok2020}.

For pure states $|\psi(\theta)\rangle$, the Fubini--Study line element is
\begin{equation}
ds_{\FS}^2=4\left(\langle d\psi|d\psi\rangle-|\langle\psi|d\psi\rangle|^2\right).
\end{equation}
This form is preferable to finite-overlap expressions because it is explicitly gauge-invariant under local phase changes of $|\psi\rangle$.

For unitary evolution generated by the Hamiltonian $\hat H$,
\begin{equation}
|\psi(t)\rangle=e^{-i\hat H t/\hbar}|\psi(0)\rangle,
\end{equation}
one obtains
\begin{equation}
ds_{\FS}=\frac{2\Delta H}{\hbar}\,dt,
\end{equation}
where
\begin{equation}
\Delta H=\sqrt{\langle\psi|\hat H^2|\psi\rangle-\langle\psi|\hat H|\psi\rangle^2}.
\end{equation}
Anandan and Aharonov emphasized the geometric character of quantum evolution and its relation to energy uncertainty \cite{AnandanAharonov1990}. In quantum estimation theory, the QFI for unitary time encoding is
\begin{equation}
F_Q(t)=\frac{4(\Delta H)^2}{\hbar^2}.
\end{equation}
Therefore,
\begin{equation}
d\LQ^2=F_Q(t)dt^2,
\end{equation}
and
\begin{equation}
d\LQ=\frac{2\Delta H}{\hbar}\,dt.
\end{equation}
The conventional interpretation reads this equation as saying that the state moves through Hilbert space as time passes. The Fisher--informational interpretation reverses the emphasis: the physically meaningful object is distinguishability in state space, while time is reconstructed by calibrating that distinguishability with a clock.

For mixed states, the infinitesimal Bures line element is related to the QFI by
\begin{equation}
ds_B^2=\frac{1}{4}F_Q(\theta)d\theta^2,
\end{equation}
with the appropriate tensor generalization in the multiparameter case. This provides a natural extension of the present construction beyond pure-state unitary dynamics.

\section{Rewriting Schr\"odinger Dynamics}
The ordinary Schr\"odinger equation is
\begin{equation}
i\hbar\frac{\partial}{\partial t}|\psi(t)\rangle=\hat H|\psi(t)\rangle.
\end{equation}
If $\LQ$ is used as the path parameter, then
\begin{equation}
\frac{\partial}{\partial t}=\frac{d\LQ}{dt}\frac{\partial}{\partial\LQ}.
\end{equation}
Using
\begin{equation}
\frac{d\LQ}{dt}=\frac{2\Delta H}{\hbar},
\end{equation}
we obtain
\begin{equation}
\frac{\partial}{\partial t}=\frac{2\Delta H}{\hbar}\frac{\partial}{\partial\LQ}.
\end{equation}
Substitution into Schr\"odinger's equation gives
\begin{equation}
i\,2\Delta H\frac{\partial}{\partial\LQ}|\psi\rangle=\hat H|\psi\rangle,
\end{equation}
or
\begin{equation}
i\frac{\partial}{\partial\LQ}|\psi\rangle=\frac{\hat H}{2\Delta H}|\psi\rangle.
\end{equation}
This expression must be interpreted as a path-dependent reparameterization rather than as a universal replacement of the Schr\"odinger equation. In particular, when $\Delta H=0$, the quantum state acquires only a global phase under a time-independent Hamiltonian and no Fubini--Study distinguishability is accumulated. In that case $\LQ$ is not a valid evolution parameter for the trajectory. More generally, when $\Delta H$ varies along the path, the generator $\hat H/(2\Delta H)$ is state-dependent through the chosen trajectory. The purpose of the expression is therefore conceptual: it shows that ordinary unitary dynamics can be represented as motion with respect to informational distance whenever the path has nonzero distinguishability speed.

\section{Gauge and Reparameterization Character of $\LF$}
The parameter $\LF$ should not be understood as a new universal external time variable. It is closer to a proper parameter on a path in state space. Given a path $\theta^i(\lambda)$, the arbitrary label $\lambda$ may be changed without altering the accumulated Fisher length. This reparameterization invariance is a desirable property: it prevents the theory from merely replacing $t$ with another arbitrary coordinate.

However, the construction also implies that $\LF$ is path- and model-dependent. In classical statistical systems, the Fisher metric depends on the chosen family of probability distributions and measurement outcomes. In quantum systems, the QFI provides a measurement-optimized metric, but it still depends on the family of states and generator of parameter changes. Therefore, $\LF$ is not an absolute cosmic clock. It is an operational measure of distinguishable change along a specified causal trajectory.

This resolves a possible misunderstanding. The proposal is not
\begin{equation}
\text{time is false and }\LF\text{ is absolute},
\end{equation}
but rather
\begin{equation}
\text{clock time is a calibration of distinguishable physical change}.
\end{equation}

\section{Relation to Quantum Speed Limits}
Quantum speed limits quantify the minimum time required for a quantum state to evolve into a distinguishable or orthogonal state. The Mandelstam--Tamm bound, for example, can be written as
\begin{equation}
\tau\geq\frac{\pi\hbar}{2\Delta H}
\end{equation}
for evolution to an orthogonal state \cite{MandelstamTamm1945}. In the present view, this relation is not fundamentally about a mysterious time--energy uncertainty. Rather, it is a bound on the rate of quantum distinguishability. Energy uncertainty controls the speed at which a state moves through projective Hilbert space.

This interpretation is aligned with modern treatments of quantum speed limits, where geometric distance and information-theoretic distinguishability play central roles \cite{Taddei2013,DeffnerLutz2013,DeffnerCampbell2017}. The novelty here is to take the informational distance not merely as a tool for bounding temporal evolution, but as a candidate precursor to clock time itself.

\section{Relation to Existing Approaches}
\subsection{Relational time and Page--Wootters}
The Page--Wootters mechanism recovers dynamics from correlations in a globally stationary state \cite{PageWootters1983}. If a clock subsystem $C$ and a system $S$ are entangled, the conditional state of $S$ relative to a clock reading can display unitary evolution. Later work by Marletto and Vedral clarified how ambiguities may be avoided \cite{MarlettoVedral2017}. The present framework agrees with the relational spirit of this approach, but emphasizes that clock readings themselves should be understood as distinguishable physical states quantified by Fisher or quantum Fisher geometry.

\subsection{Thermal time}
The thermal time hypothesis proposes that the physical flow of time is determined by the statistical state of a generally covariant quantum system, using modular flow from the algebraic structure of observables \cite{ConnesRovelli1994,Rovelli2009ThermalTime}. The Fisher--informational approach is similar in rejecting universal external time, but differs in its proposed primitive: rather than modular thermal flow, it uses causal distinguishability measured by Fisher geometry.

\subsection{Problem of time in quantum gravity}
The problem of time arises because time plays incompatible roles in general relativity and quantum mechanics \cite{Isham1993,Kuchar2011,Anderson2012}. A Fisher--informational parameter could provide an internal ordering variable that does not presuppose an external background time. This paper does not solve quantum gravity, but it offers a language in which evolution is not fundamental temporal flow but distinguishability along causally admissible state trajectories.

\subsection{Information geometry and stochastic dynamics}
The connection between Fisher information, statistical length, and dynamical evolution has also appeared in stochastic thermodynamics and nonequilibrium physics. Ito's work, for example, relates the time evolution of stochastic averages to Fisher information and statistical length \cite{Ito2020}. These results support the broader idea that dynamical change can be characterized geometrically through information measures. The present work differs by interpreting such distinguishability as a candidate precursor to clock time itself.

\section{Quantitative Examples}
\subsection{Example I: Classical oscillator as phase clock}
Consider a harmonic oscillator
\begin{equation}
x(t)=A\cos(\omega t+\phi_0).
\end{equation}
The directly relevant internal variable is phase,
\begin{equation}
\phi=\omega t+\phi_0.
\end{equation}
Thus,
\begin{equation}
x(\phi)=A\cos\phi,
\end{equation}
and conventional time is reconstructed as
\begin{equation}
t=\frac{\phi-\phi_0}{\omega}.
\end{equation}
A clock is precisely a system whose phase is used as a reference. This simple model illustrates that time can be treated as a calibration of physical state progression.

\subsection{Example II: Qubit clock and QFI}
Let
\begin{equation}
\hat H=\frac{\hbar\omega}{2}\sigma_z,
\end{equation}
and
\begin{equation}
|\psi(0)\rangle=\frac{1}{\sqrt{2}}\left(|0\rangle+|1\rangle\right).
\end{equation}
Then
\begin{equation}
|\psi(t)\rangle=\frac{1}{\sqrt{2}}\left(e^{-i\omega t/2}|0\rangle+e^{i\omega t/2}|1\rangle\right).
\end{equation}
The overlap with the initial state is
\begin{equation}
|\langle\psi(0)|\psi(t)\rangle|^2=\cos^2\left(\frac{\omega t}{2}\right).
\end{equation}
Here
\begin{equation}
\Delta H=\frac{\hbar\omega}{2},
\end{equation}
so
\begin{equation}
F_Q(t)=\omega^2,
\end{equation}
and
\begin{equation}
\LQ=\int_0^t\sqrt{F_Q(t')}\,dt'=\omega t.
\end{equation}
Thus, for this ideal qubit clock,
\begin{equation}
t=\frac{\LQ}{\omega}.
\end{equation}
Clock time is reconstructed from accumulated quantum distinguishability calibrated by the angular frequency $\omega$. The state does not need an ontological time substance to acquire operational meaning; it needs a distinguishable trajectory and a frequency calibration.

\subsection{Example III: Exponential decay as internal distinguishability}
The standard decay law is
\begin{equation}
N(t)=N_0e^{-\Gamma t}.
\end{equation}
Define an internal decay parameter
\begin{equation}
\Lambda_D=-\ln\left(\frac{N}{N_0}\right).
\end{equation}
Then
\begin{equation}
N(\Lambda_D)=N_0e^{-\Lambda_D},
\end{equation}
and ordinary time is reconstructed as
\begin{equation}
t=\frac{\Lambda_D}{\Gamma}.
\end{equation}
This example shows that temporal laws can often be rewritten in terms of process-internal change variables. A more statistical version can be obtained from the survival probability distribution, where Fisher information quantifies the precision with which the decay parameter can be inferred from observed events.

\subsection{Example IV: Clock quality as Fisher distinguishability per tick}
Let a clock produce a sequence of probability distributions $p(x|N)$ associated with tick number $N$. Define the Fisher distinguishability per tick as
\begin{equation}
\eta_C=\frac{d\LF^{(C)}}{dN}.
\end{equation}
A useful clock is not merely a high-frequency system; it is a system whose successive states are stable, reproducible, and distinguishable. One may therefore define a Fisher stability functional
\begin{equation}
\mathcal{S}_C=\left[\mathrm{Var}\left(\frac{d\LF^{(C)}}{dN}\right)\right]^{-1},
\end{equation}
which measures the regularity of distinguishability accumulation per tick. This points toward a metrological interpretation of clock quality based on Fisher geometry rather than an assumed external time variable.

\subsection{Example V: Cosmological redshift and the universal photosphere}
Cosmology routinely uses redshift $z$ or scale factor $a$ as an evolution parameter:
\begin{equation}
1+z=\frac{1}{a}.
\end{equation}
Many observables are expressed as functions of $z$ rather than time. In a Fisher--informational cosmological formulation, one could define
\begin{equation}
d\Lambda_{\mathrm{cos}}^2=F^{\mathrm{cos}}_{ij}d\theta^i d\theta^j,
\end{equation}
where $F^{\mathrm{cos}}_{ij}$ is the Fisher matrix for cosmological parameters inferred from CMB anisotropies, large-scale structure, supernovae, or other observations.

This connects naturally with the interpretation of the cosmic microwave background last-scattering surface as a finite-thickness optical layer rather than a material boundary. Following the terminology proposed here, the last-scattering surface may be described as a \emph{universal photosphere}: an observer-dependent electromagnetic information surface defined by photon decoupling.

\section{Conceptual Structure}
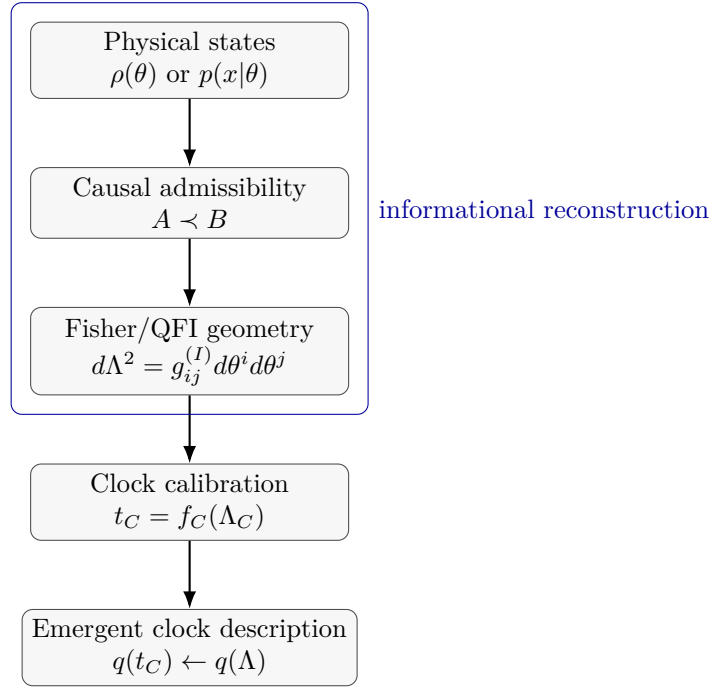
\begin{figure}[t]
\centering
\begin{tikzpicture}[node distance=0.9cm, every node/.style={font=\small}, box/.style={rectangle, rounded corners, draw=black!70, align=center, minimum width=4.2cm, minimum height=0.9cm, fill=black!3}, arrow/.style={-Latex, thick}]
\node[box] (states) {Physical states\\$\rho(\theta)$ or $p(x|\theta)$};
\node[box, below=of states] (causal) {Causal admissibility\\$A\prec B$};
\node[box, below=of causal] (fisher) {Fisher/QFI geometry\\$d\Lambda^2=g^{(I)}_{ij}d\theta^i d\theta^j$};
\node[box, below=of fisher] (calib) {Clock calibration\\$t_C=f_C(\Lambda_C)$};
\node[box, below=of calib] (time) {Emergent clock description\\$q(t_C)\leftarrow q(\Lambda)$};
\draw[arrow] (states) -- (causal);
\draw[arrow] (causal) -- (fisher);
\draw[arrow] (fisher) -- (calib);
\draw[arrow] (calib) -- (time);
\node[draw=blue!60!black, rounded corners, fit=(states)(causal)(fisher), inner sep=0.25cm, label={[blue!60!black]right:informational reconstruction}] {};
\end{tikzpicture}
\caption{Conceptual structure of Fisher--informational time. Physical states acquire operational meaning through causal order and distinguishability. Fisher or quantum Fisher geometry defines an accumulated informational parameter $\Lambda$. Ordinary clock time appears only after calibration by a stable reference process.}
\label{fig:concept}
\end{figure}

\section{Potential Experimental and Theoretical Program}
The proposal becomes scientifically useful only if it generates calculations. We identify several possible directions.

\subsection{Fisher clocks}
A clock may be characterized by the Fisher distinguishability gained per cycle:
\begin{equation}
\eta_C=\frac{d\LF^{(C)}}{dN}.
\end{equation}
Clock quality could then be measured not merely by frequency stability, but by distinguishability stability. This may connect the foundations of time with quantum metrology and atomic-clock precision \cite{Giovannetti2006QuantumMetrology,Pezze2018QuantumMetrology}.

\subsection{Open quantum systems}
In open systems, decoherence produces records and preferred bases. One may define a record-based informational parameter
\begin{equation}
\Lambda_{\mathrm{rec}}=\sum_k D(\rho_k\|\rho_{k-1}),
\end{equation}
where $D$ is a distinguishability measure such as relative entropy. The arrow of time could then be associated with monotonic growth of accessible records rather than a primitive temporal flow.

\subsection{Cosmological Fisher time}
Given a set of cosmological observables $\mathcal{O}$, the Fisher matrix
\begin{equation}
F_{ij}=\left\langle \frac{\partial\ln\mathcal{L}}{\partial\theta^i}\frac{\partial\ln\mathcal{L}}{\partial\theta^j}\right\rangle
\end{equation}
can define an informational distance between cosmological states. This suggests a possible reconstruction of cosmic evolution from observational distinguishability rather than from cosmic time alone.

\subsection{Quantum speed-limit reinterpretation}
Quantum speed limits already connect distinguishability, energy uncertainty, and minimal evolution times. In the present theory, these results are reinterpreted as evidence that the fundamental measurable quantity is the distinguishability rate, while time is a calibrated parameter.

\section{Limitations}
Several limitations must be acknowledged.

First, the proposal is not yet a complete replacement for relativistic or quantum dynamics. It is a reformulation program. Second, Fisher information depends on the statistical model and measurement context, while QFI depends on the chosen family of quantum states and generators. A fully fundamental theory must specify which information metric is physically primary. Third, the relation between causal order and Fisher geometry requires further development, especially in relativistic and quantum-gravitational settings. Fourth, the framework must identify cases in which it yields predictions or conceptual advantages not obtainable from standard time parameterizations. Fifth, $\LF$ can fail to be monotonic or well-defined for trajectories that do not accumulate distinguishability, such as stationary energy eigenstates in closed systems.

These limitations are important. They prevent the proposal from being overstated. The current objective is to formulate a mathematically precise and physically motivated research program.

\section{Originality Statement}
Relational, thermal, and emergent views of time are not new. The originality of the present work does not lie in the general claim that time may be relational or emergent. Rather, it lies in proposing a Fisher-geometric causal parameter, $\LF$, as an operational precursor of clock time. In this framework, clocks do not reveal an external temporal substance; they calibrate accumulated distinguishability along causally admissible trajectories of physical states.

Thus, the proposed contribution is not the general idea that time is emergent, but the use of Fisher information and quantum Fisher information as operational metrics from which clock time is reconstructed.

\section{Conclusion}
We have proposed a Fisher--informational interpretation of time. The central claim is that clocks do not measure time as an independent substance. They instantiate reproducible physical processes whose distinguishable states can be correlated with other events. The proposed parameter $\LF$ quantifies accumulated Fisher distinguishability along causally ordered trajectories in state space. Ordinary clock time is then reconstructed as a calibration of this informational parameter.

The proposal is compatible with relational time, Page--Wootters dynamics, thermal time, quantum speed limits, information geometry, and the problem of time in quantum gravity, but offers a distinct emphasis: Fisher distinguishability is treated as a candidate precursor to clock time. The key statement is:
\begin{quote}
\emph{Time is not measured by clocks; clock time is reconstructed from the Fisher distinguishability accumulated along causally ordered physical changes.}
\end{quote}
This provides a concrete research direction for a causal--informational physics in which temporal evolution is emergent from distinguishability, correlation, and causal order.

\section*{Acknowledgments}
The author acknowledges discussions on the operational, relational, and informational interpretation of time. This draft is intended as a conceptual foundation for further mathematical and physical development.

\section*{Disclosure Statement}
The author declares no conflicts of interest.

\section*{Data Availability Statement}
No data were used in this theoretical study.

\end{document}